\documentclass[10pt,letterpaper]{article}
\usepackage{opex3}

\begin{document}

\title{Experimental Single-Photon Transmission from Satellite to Earth}

\author{Juan Yin$^{1,3}$, Yuan Cao$^1$, Shu-Bin Liu$^1$, Ge-Sheng Pan$^1$, Jin-Hong Wang$^1$, Tao Yang$^1$, Zhong-Ping Zhang$^2$, Fu-Min Yang$^2$, Yu-Ao Chen$^1$, Cheng-Zhi Peng$^{1,4}$, and Jian-Wei Pan$^1$}

\address{\textit{$^1$}Shanghai Branch, National Laboratory for Physical Sciences at Microscale 
\\and Department of Modern Physics, University of Science and Technology \\of China, 
Shanghai 201315, China\\
\textit{$^2$}Shanghai Astronomical Observatory, Chinese Academy of Sciences, \\Shanghai 200030, China\\}

\email{\textit{$^3$}yinjuan@ustc.edu.cn} 
\email{\textit{$^4$}pcz@ustc.edu.cn} 



\begin{abstract}
Free-space quantum communication with satellites opens a promising avenue for global secure quantum network and large-scale test of quantum foundations. Recently, numerous experimental efforts have been carried out towards this ambitious goal. However, one essential step - transmitting single photons from the satellite to the ground with high signal-to-noise ratio (SNR) at realistic environments - remains experimental challenging. 
Here, we report a direct experimental demonstration of the satellite-ground transmission of a quasi-single-photon source. In the experiment, single photons($\sim$0.85~photon per pulse) are generated by reflecting weak laser pulses back to earth with a cube-corner retro-reflector on the satellite Champ, collected by a 600-mm diameter telescope at the ground station, and finally detected by single-photon counting modules (SPCMs) after 400-km free-space link transmission. With the help of high accuracy time synchronization, narrow receiver field-of-view (FOV) and high-repetition-rate pulses (76~MHz), a SNR of better than 16:1 is obtained, which is sufficient for a secure quantum key distribution. Our experimental results represent an important step towards satellite-ground quantum communication. 
\end{abstract}




\section{Introduction}
Quantum communication is proven to be only unconditionally secure method for information exchange, which could well be the first commercial application of quantum information science. Due to the low channel loss, free-space optical channels become far superior over fiber links to achieve ultra-long-distance quantum communication. With the help of satellite, it is possible to realize quantum communication networks and tests of quantum foundations on a global scale. Significant experimental efforts have been devoted to investigate the feasibilities of satellite-based quantum communications. With the development of technologies, a series of researches have extended the communication distance of quantum key distribution (QKD)\cite{Schmitt-Manderbach07}, entanglement distribution\cite{Yin12}, and quantum teleportation\cite{Yin12,Ma12} between fixed locations to 100-km scale. Furthermore, full-scale verifications of ground-satellite QKD have been reported by utilizing a moving platform on a turntable and a floating platform on a hot-air balloon\cite{Wang13}. 

In the meantime, besides the experimental efforts in the ground, satellite based quantum communication projects have been proposed by several different countries\cite{Xin11,QUEST,Jennewein13} and the launching time has been scheduled. Before the launching of the satellite, however, a direct study of the whole process of transmitting and detecting single photons from the satellite to the ground with the help of existing satellite or aircraft at realistic envirments is necessary but remains experimental challenging owing to the high background noise level from all stars and the difficulties of tracking and synchronization. 

 
Along this direction, an EU joint group reported an interesting attempt to verify the efficiency of atmospheric transmission\cite{Villoresi08} by sending laser pulses from the ground station to the satellite Ajisai equipped with cube-corner retroreflectors and further detecting the reflected laser signal with a modified classical laser ranging system. However, the simulated ``transmitting'' source on the satellite contains more than 1000 photons per pulse, which is far beyond single-photon level as the prerequisite of quantum communication. Further more, even with such a source and the time bin of 5~ns used in the experiment\cite{Villoresi08} for analyzing data, the obtained SNR is less than 1:1, which is not sufficient to create any secure key. Thus such a study failed to address the problem that whether it is possible to create satellite-ground quantum channel in the presence of all stray light. Here, we report for the first time an experimental simulation of a quasi-single-photon transmitter on the satellite with an average photon number of 0.85 per pulse and a full divergence angle of 38~$\mu$rad sending to the ground. In the ground station, by utilizing the coarse tracking techniques developed in Ref.~\cite{Wang13}, taking advantage of high-repetition-rate pulses (76~MHz) and improving the synchronization accuracy, we succeed to detect the desired single photons passing through the 400-km free-space channel with a SNR of 16:1. Such a SNR is good enough to generate quantum links for unconditionally secure QKD\cite{Wang13}. Hence, our results represent an important step towards satellite-to-ground QKD.

\section{Scheme}

\begin{figure}[Fig_Scheme]
\begin{center}
\includegraphics[width=7cm]{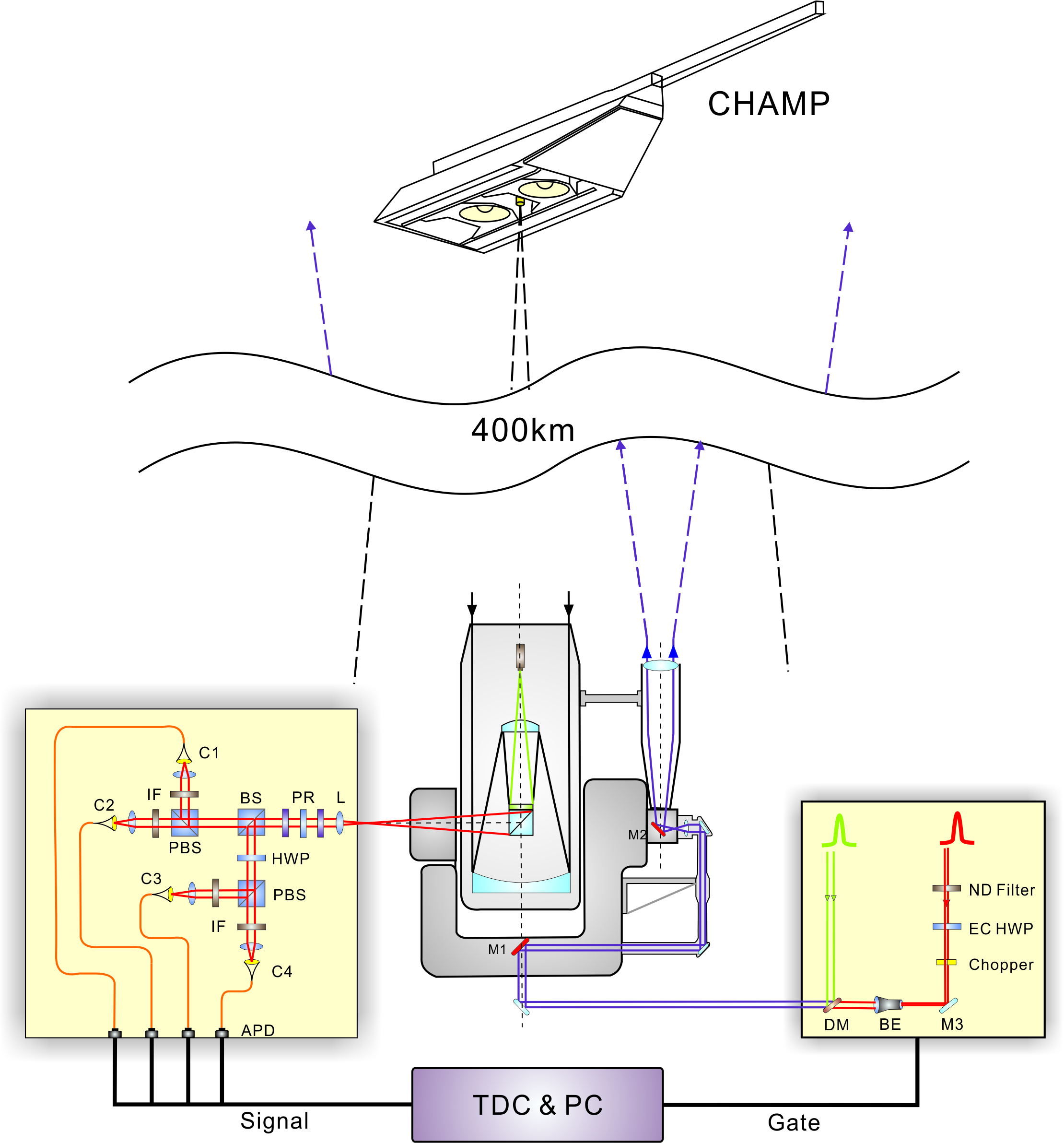}
\caption{The scheme of the single-photon link from satellite to experimental setup installed at the Shanghai Observatory. CHAllenging Minisatellite Payload (Champ) was a German satellite launched July 15th, 2000 from Plesetsk, Russia and was used for atmospheric and ionospheric research, as well as other geoscientific applications. It was covered by 2~cm-diameter retroreflectors on each side. A train of pulses of 3~ps duration, 702~nm wavelength, 0.4~nJ of energy and 76~MHz repetition rate are coupled with the laser ranging pulses before being sent toward the transmitting telescope about 20~cm in aperture. The neutral density (ND) filter is used to control the weak pulse light, and the chopper is used to filter the backscattering noise. After the beam spreading, a fraction of the beam in the uplink path irradiates the satellite Champ. The corner cubes on the satellite retro-reflect back to the Earth a small portion of the photons in the laser pulse (downlink), which is the single-photon channel. Some of the photons in the downlink path are collected by the receiving telescope, a reflecting telescope with aperture of 60 cm, and detected by SPCMs, placed behind polarization measurement devices and spectral filters. The transmitting telescope and the receiving telescope are separated by a distance of 30~cm. All relevant events in the time domain are recorded by time measurement system TDC, and then referred to as coordinated universal time (UTC).\label{Fig_Scheme}}
\end{center}
\end{figure}

As shown in Fig.~\ref{Fig_Scheme}, our experimental system consisted of a tracking and pointing telescope, a high repetition rate laser, a receiving terminal for collecting and detecting single-photons and a data-acquisition system. The telescope employed the binocular structure which could achieve the separation of emitting and receiving photons. The transmitting and receiving telescopes, with diameter of 20 cm and 60 cm respectively, are separated by 30~cm. The main advantage of this structure is that we can modulate the transmitting optical signal, and provide multiple-measurement basis on quantum signal conveniently. On the other hand, in order to realize the independence and synchronization between detecting quantum signal and laser ranging signal, we located dichroic mirrors (DMs) in both transmitting and receiving optical path.

At the transmitting side, a laser beam (wavelength of 702~nm, repetition rate of 76~MHz, single-pulse energy of 0.4~nJ, and pulse width of 3~ps) was coupled by a dichroic mirror with a green laser beam from the laser ranging system of Shanghai Observatory. Then the coupled beams were transmitted from the transmitting telescope. Due to employment of a laser with high repetition rate, it made a possibility that we can receive enough single-photons coming from the satellites in a few minutes. The neutral density (ND) filter in the transmitting path was used to achieve energy control of the weak light pulse. The chopper was used to wipe out backscattering noise. The electronically controlled half-wave plate (EC HWP) was a controller of real-time polarization tracking based on the kinematical reference system. 

At the receiving side, an optical detection was installed in the place of an original complementary metal-oxide semiconductor (CMOS). After the light reflected by the satellite enters the receiving telescope, it was focused by a primary and a secondary mirror. A meniscus lens reflected the light whose wavelength was 532 nm to the ocular. Because of the perfect separation of quantum signal and laser ranging signal, we could operate the complete polarization analysis of the incoming photons which will detected by SPCMs. The transmitting and receiving photoelectric conversion signal would be led into a time-to-digital converter (TDC) and analyzed by a computer.

\begin{figure}[Fig_Scheme]
\begin{center}
\includegraphics[width=7cm]{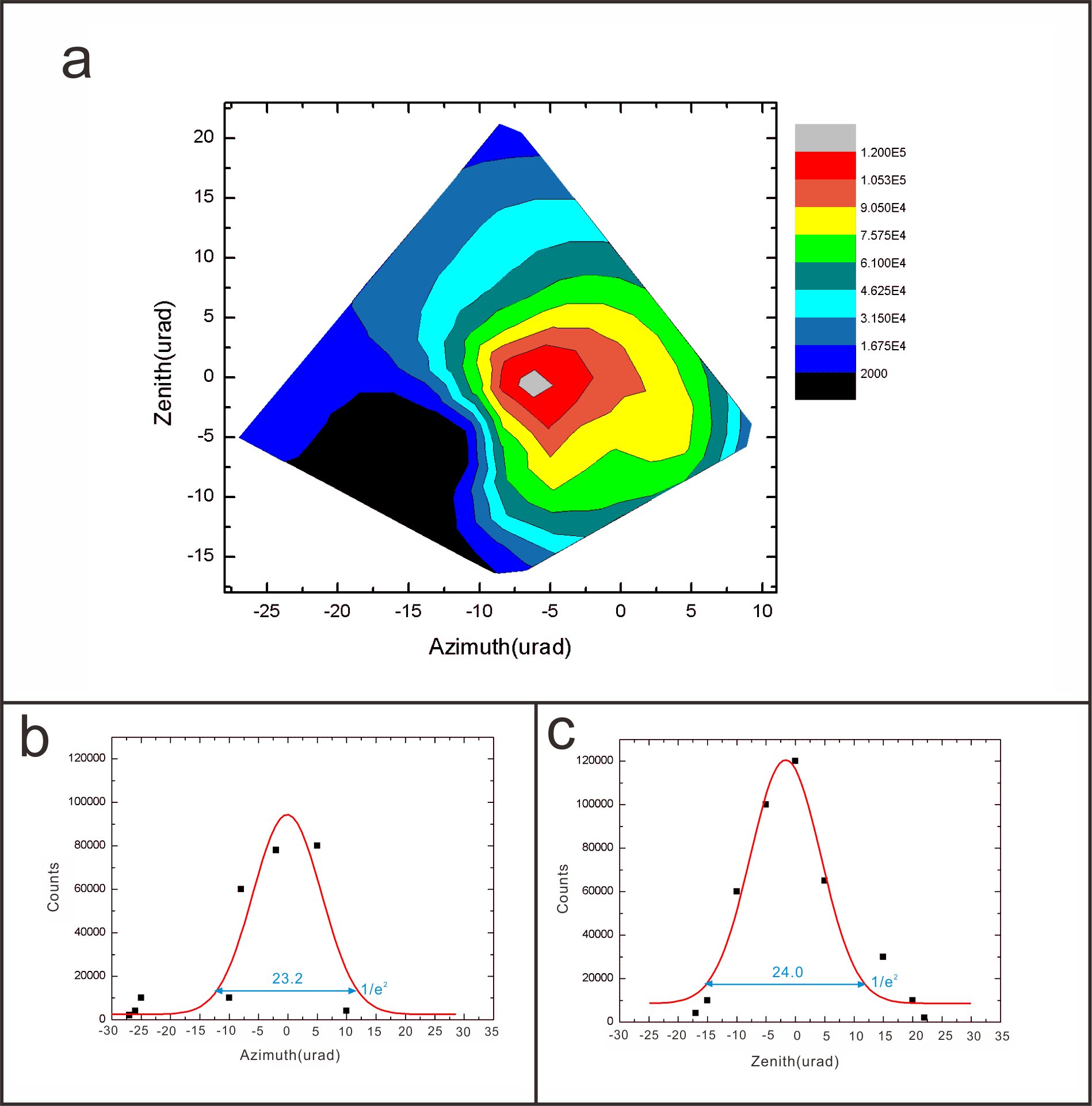}
\caption{FOV of the detection system testing by scanning and recording the light intensity of the Polaris. The telescope was tracked to the Polaris, which could be seen as a fixed star. By changing the relative telescope tracking point, we record the different count rates of the changing position by SPCMs. Fig.2a was the three-dimensional draw according to the counts. Fig. 2b was the fitting curve of FOV in the azimuth axial.  Fig. 2b was the fitting curve of FOV in the zenith axial.The FOV of the receiving system, in both the azimuth and zenith axials, were measured to be less than 5$''$. \label{Fig_View}}
\end{center}
\end{figure}

In this experiment, we have made some improvements to reduce background noise, such as adopting spatial, spectral and time filtering.
We test the receiving FOV of the detection system by scanning the telescope and recording the light intensity of the Polaris. As shown in Fig.~\ref{Fig_View}, we experimentally demonstrated that FOV is 5$''$. Besides, a narrow band filter with the bandwidth of 3~nm  was added to the receiving optical path to filter out background noise (including reflected lights from the sky and the ground) from the desired signal. In addition, we sectioned the transmitted pulsed laser from time sequence and filtered the the backscattering by controlling the detection time. Every time after a transmitted laser pulse had passed the aerosphere,  determining by the influence of the backscattering from the aerosphere was obviously within 30 km (round trip time of 200~$\mu$s), we detected the echo signal and avoided these kinds of noise. Besides, we did our experiment at night and choose the satellite in the shadow of the earth. Thus, even on a clear night we manage to reduce the background noise above the zenith angle of 45$^0$ to be less than 100~counts per second (cps). 

The key experimental parameters were shown in table 1. Here, the pulse energy $E$ was 0.4~nJ (702~nm); $S$, the photon number per joule, was $3.53\times 10^{18}$ (702 nm); $A_s$, the effective area of satellite corner retro-reflector, was 11.34 $\textit{cm}^2$ for Champ; $A_r$, the effective area of receiving telescope, was 0.25 $\textit{m}^2$ ; $K_t$ was the efficiency of transmission system of 0.2; $K_r$ was the efficiency of optical receiving system of 0.15, including 0.5 for receiving telescope, 0.5 for coherent filter and 0.6 for multimode fiber collection; $T$ was one-way atmospheric transmission of 0.6; $\eta$ was the quantum efficiency of receiving detection device of 0.65; $\alpha$ was the attenuation factor (including the influence of satellite retro-reflector efficiency, atmospheric jitter and turbulence) of 13 dB; $R$ was the satellite distance of 400 km for Champ; $\theta_t$ was the laser beam divergence angle of 300 $\mu$rad; $\theta_s$ was the diffraction angle of satellite retro-reflector of 38~$\mu$rad for Champ\cite{Neubert98}.

\begin{table}
\centering\caption{Parameters of satellite-to-ground quantum channel link.}
\begin{tabular}{c | c}
  \hline
  Items & Parameter \\ \hline
  Laser repetition rate $\Lambda$ & 76MHz \\
 Single pulse energy $E$ & 0.4nJ \\
 Photon number per joule $S$ & 3.53$\times 10^{18}$ \\
 Champ satellite height $R$ &  400km \\
 Geometrical efficiency of transmission $\frac{4A_s}{\pi\theta^2_tR^2}$ & 1.00$\times 10^{-7}$ \\
 Geometrical efficiency  of receiving $\frac{4A_r}{\pi\theta^2_sR^2}$ & 1.38$\times 10^{-3}$ \\
 Efficiency of transmission system $K_t$ & 0.20 \\
 Receiving system efficiency $K_r$ & 0.15 \\
 Quantum detection efficiency $\eta$ & 0.65 \\
 One-way atmospheric transmission $T$ & 0.60 \\
 Attenuation factor $\alpha$ & 0.05 \\
  \hline
\end{tabular}
\end{table}

According to the parameters in Table 1, we could calculate the one-way channel attenuation from ground to satellite:
\begin{equation}
{\eta_{up}= \frac {4\cdot A_s\cdot K_t\cdot T\cdot \alpha} {\pi\cdot R^2\cdot \theta^2_t}=6.00\times 10^{-10}}.
\end{equation} 
The numbers of photons reflected from satellite was:
\begin{equation}
{N_0=E\cdot S\cdot \eta_{up}=0.85}.
\end{equation} 
Attenuation in the one-way channel from satellite to earth was:
\begin{equation}
{\eta_{down}= \frac {4\cdot A_r\cdot K_r\cdot T\cdot \eta} {\pi\cdot R^2\cdot \theta^2_s}=8.07\times 10^{-5}}.
\end{equation} 
Therefore, the theoretical number of photons detected by each pulse at the receiving end was:
\begin{equation}
{N=N_0\cdot \eta_{down}=6.83\times 10^{-5}}.
\end{equation} 

\section{Result}
In our experiment, the effective measurement time of the return signal of the Champ satellite (400~km) in each period ( $\tau=16~\textit{ms}$) of the chopped wave was $\tau_0=1.65~\textit{ms}$ . For example, the number of echo signals in 10~s was 5385. Thus the average counts per second were  $\overline{N}=5222$. The measured dark counts per second were$\overline{N_b}=89$. Therefore, the photon number per echo generated by a single laser pulse was experimentally measured:

\begin{equation}
{N_{exp}= \frac {\overline{N}-\overline{N_b} } {\Lambda}=6.75\times 10^{-5}}
\end{equation} 

The repetition rate of the present traditional laser ranging system (LRS) could be achieved at the magnitude of kHz. However, a light source having a higher repetition rate and time resolution was required due to the high loss in the channel. In the time synchronization system of our experiment, the quantum system and the laser ranging system were independent. By this way we got rid of the frequency limitation of the laser ranging system. We also utilized a TDC with high accuracy to record the arrival time of the photons and sample the pulsed photons\cite{Song06}. After the photons were reflected by the satellite, passed through a 400~km satellite-to-ground link and the atmosphere, they would be detected. We defined $t_0$ as the emission time and $t$ as the corresponding return time. According to the return time $t$ of each photon, and the real-time satellite distance $t_c$ , we could calculate the theoretical transmitting time $t_{\textit{exp}}$ , We used the difference $D=t_0-t_{\textit{exp}}$ to characterize the accuracy of the time information.

\begin{figure}[Fig_Champ]
\begin{center}
\includegraphics[width=7cm]{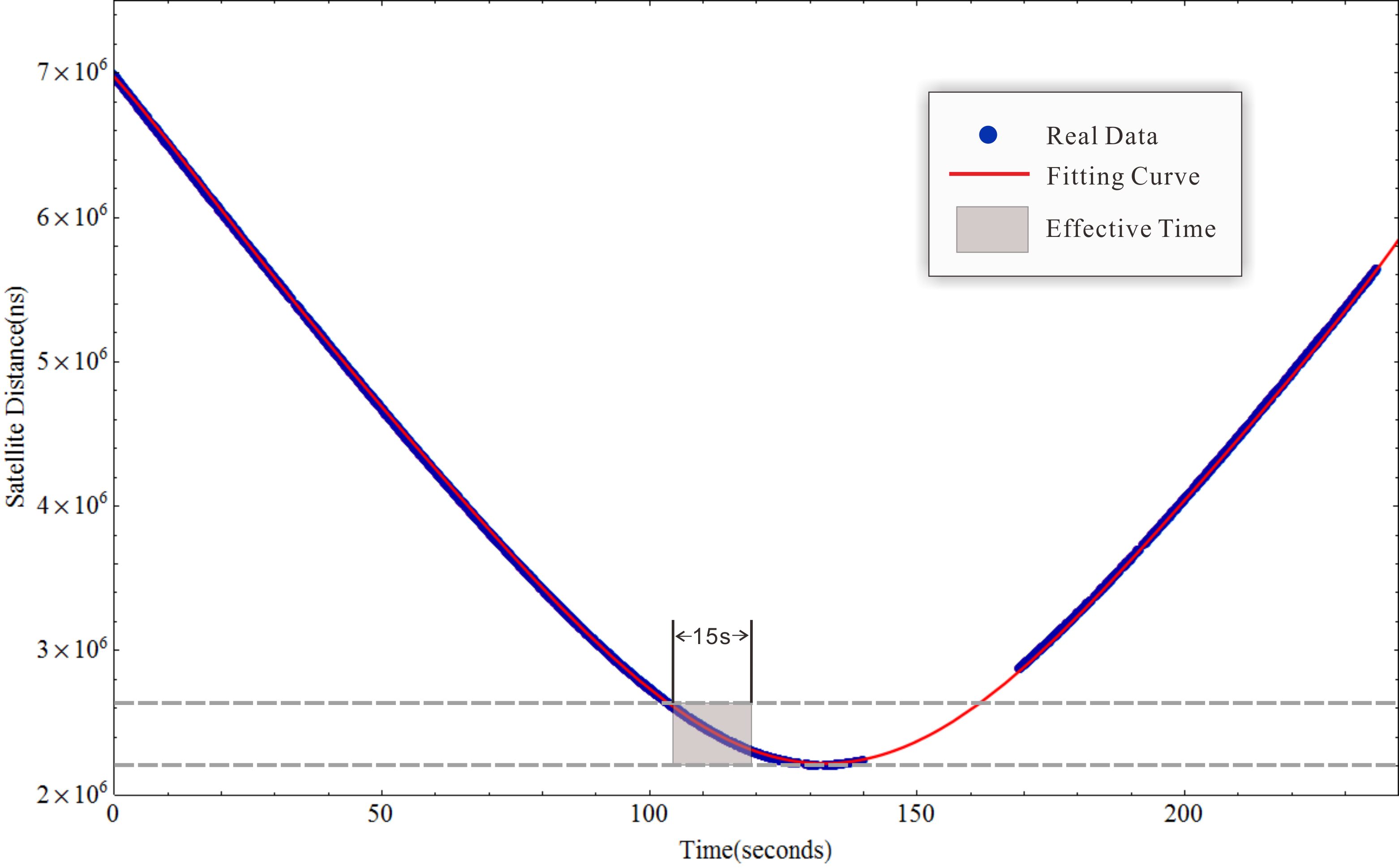}
\caption{Plot of the range Rs between the satellite Champ and Shanghai Observatory. The round-trip time of the signal, directly proportional to $R_s$, exhibits strong and rapid variations during the satellite pass.  The blue points are the LSR system's records and the red line is the fitting curve on the survey data. The efficient time with maximal counts of single photons reflected from Champ and detected by SPCMs is shown in the gray box. The perigee height of Champ satellite that day was about  330km. The round-trip was ranging from 2.63~ms to 2.3~ms during our records of 15~s. \label{Fig_Champ}}
\end{center}
\end{figure}

\begin{figure}[t!]
\begin{center}
\includegraphics[width=7cm]{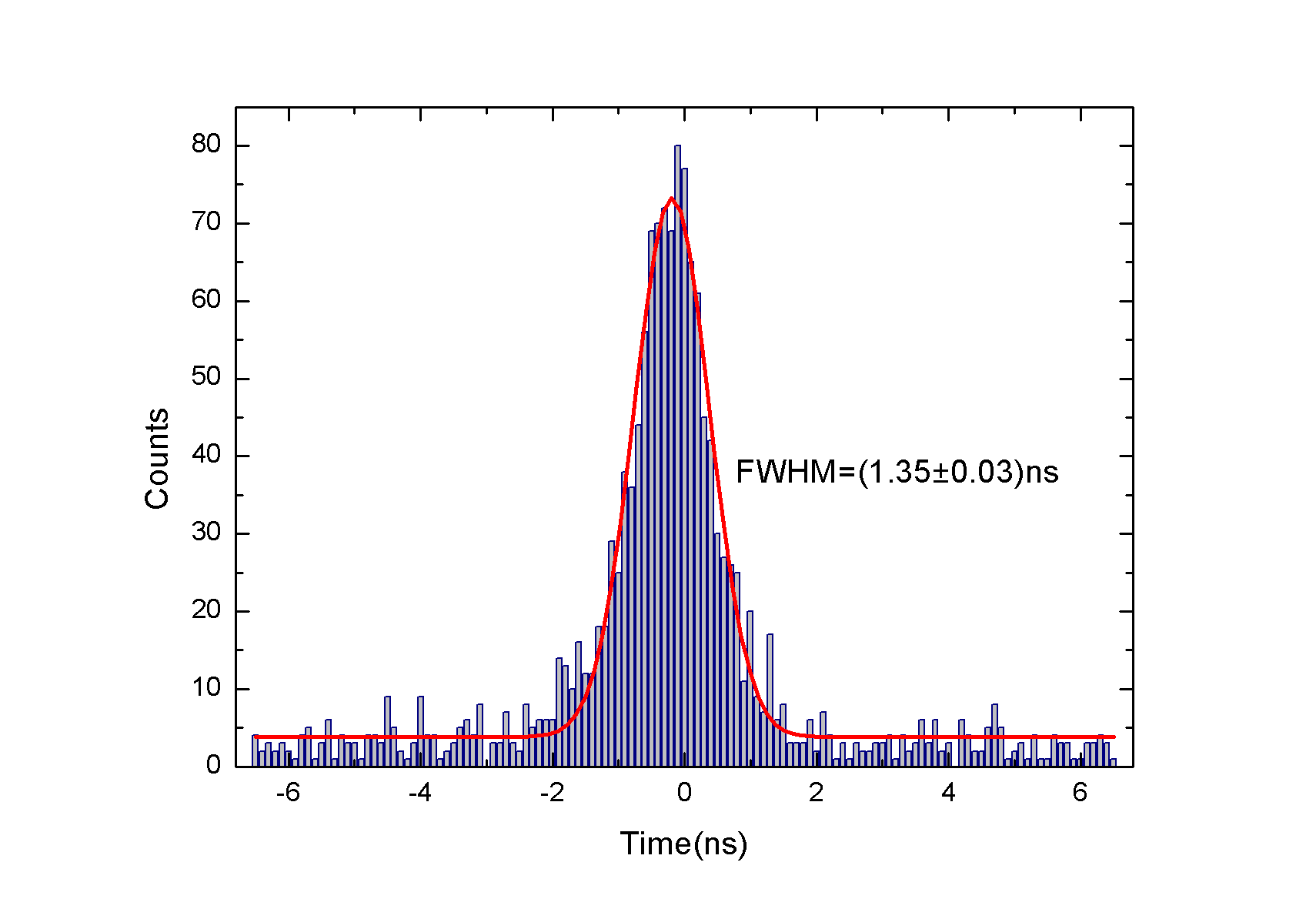}
\caption{Histogram of all \textit{D} values between expected and observed detections for Champ satellite.We summarized the D values numerically with 0.1ns as the unit. The peak of the histogram is centered at 0~ns, as expected. By Gaussian fitting, the full width at half maximum (FWHM) time accuracy was observed $1.35~ns\pm0.03~ns$.
\label{Fig_Space}}
\end{center}
\end{figure}

Although the photons were emitted in pulsed intervals, the return time which we detected was disordered due to the changing distance of the satellite. To identify which pulse the photons belong to, we had to take the satellite distance measured by the laser as a synchronization data and obtained the round-trip time deviation by performing of offline analysis. The satellite orbit fitting curve and histogram of all D values with 0.1 ns as the unit were shown in Fig.~\ref{Fig_Space}. The time synchronization accuracy was $1.35\pm0.03~\textit{ns}$ at the full width at half maximum (FWHM). The main factors influencing time accuracy included: 350~ps for the time resolution of SPCMs, 160~ps for the accuracy of TDC, 1000~ps for the fitting precision of satellite orbit, and the rest effect for the time jitter of our pico-second pulsed laser.

By using 2 ns bin size to deal with the experimental data, we got the effective echo signals counts $\overline{N'}=1000$ and the effective dark counts  $\overline{N'_b}=58$. The SNR was calculated as follow:

\begin{equation}
{SNR= \frac {\overline{N'}-\overline{N'_b} } {\overline{N'_b}}=16.2}
\end{equation} 

\section{Conclusion and perspectives}
In summary, our experiment demonstrated the direct simulation of the single photon transmission through a satellite-to-ground free-space channel. From the quasi-single-photon transmitter on the satellite Champ, we observed the desired single photons with a counting rate up to 570~cps, a SNR of better than 16:1 and a time accuracy of 1.35~ns. These results are sufficient to set up an unconditionally secure QKD link between satellite and earth, technically\cite{Wang13}. In the scheduled Chinese Quantum Science Satellite\cite{Xin11}, both the brightness of the light source and the divergence angle of the transmitter will be well improved and the fine acquiring, pointing and tracking techniques will be employed\cite{Yin12,Wang13}. Together with the field tests\cite{Yin12,Wang13}, our results represent an crucial step towards the final implementation of high-speed QKD between the satellite and the ground stations, which will also serve as a test bed for secure intercontinental quantum communication. 

~\par
\noindent \textbf{Acknowledgement}
~\par
~\par
\noindent We acknowledge insightful discussions with Qiang Zhang.
This work has been supported by the NNSF of China, the CAS, the National Fundamental Research Program (under Grant No.  2011CB921300) and NSERC.

\end{document}